# Comment on "Orientational Distribution of Free O-H Groups of Interfacial Water is Exponential"

In a recent letter [1], Sun et al. reported that combined MD simulation and sum frequency generation vibrational spectroscopy (SFG-VS) measurements led to conclusions of a broad and exponentially decaying orientational distribution, and the presence of the free O-H group pointing down to the bulk at the air/water interface. In this comment, we show that their main conclusions are based on questionable interpretation of the SFG-VS data presented in the letter [1], and are also contrary to the established data analysis and interpretations in the literature [2-5].

The idea and systematic analysis of multiple experimental geometries, as the basis of experimental study in Ref. [1], to investigate the SFG-VS spectra and to analyze the orientational distribution of the free O-H group at the air/water interface was first demonstrated and established in 2006 by Gan et. al [2]. For the four geometries with the visible incident angle from 37° to 63°, the free O-H peak intensity in the ssp polarization configuration remained nearly the same, while that in the ppp changed strikingly two orders of magnitude. This can only be interpreted with a tilt angle from the surface normal with a narrow orientational distribution regardless of the form of distribution function [2,3], and an upper limit of $\sigma = 15°$ distribution width was estimated from the data with a Gaussian distribution function [2]. Recent study on the free O-H SFG-VS at the air/water interface of the NaF solution with different NaF bulk concentration put additional unquestionable affirmation of the narrowness of the free O-H distribution [5]. As the NaF bulk concentration increase from 0 to 0.94M, the ppp intensity of the free O-H peak was reduced to zero, while ssp intensity remained the same. Only narrow tilt angle orientational distributions for the free O-H group at the air/water interface can possibly result in zero values in the ppp intensity while the ssp intensity remains significant. Even though this work on free O-H at NaF aqueous solution surface (Ref. [5]) was cited several times in Ref. [1], indicating strong relevance between the two works, it nevertheless appears to us that the significance and the relevance in the analysis and conclusion in Ref. [5] were not properly understood by the authors of Ref. [1].

The SFG-VS data in the Ref. [1], obtained with two geometries, were consistent with the previous reported data [2,3,5], even though with less ideal quality. For example, it has been well established that the ssp spectra in different geometries should have identical spectral line shape in the whole spectral region [2,4,5]. Even though it has been shown that the ssp spectra from different groups overlapped nicely with each other in surface SFG spectra, the ssp spectra in Ref. [1] (Fig. 2a and 2d) clearly did not. Nevertheless, the ppp intensities of the free O-H peak in the two geometries (Fig. 2a and 2d) did differ by at least one order of magnitude and are consistent with prior measurements in literature [2,4]. Such change of the ppp intensity of the free O-H peak in the experiment cannot be accounted in the orientational curves in the Fig. 3b and 3c, where the ppp amplitude are essentially close to each other, even with the change of the polarizability ratio $r$ changed from 0.32 to 0.15, and with the apparent tilt angle changed from about 30°-40° to 63° [1].

In fact, the broad orientational distribution as in Ref. [1], or any distribution as broad, cannot interpret the significant intensity change in the ppp polarization of the free O-H peak, while ssp intensity remains nearly the same, as in the data in Ref. [1] and in the literature [2-4]. The disappearance of zero intensity crossing point in the simulated ppp orientational curves in Fig. 3b and 3c in Ref. [1] is the direct result of the so-claimed "broad, exponentially decaying distribution". Such zero-crossing point is the key signature of the $C_{\infty v}$ symmetry of the free O-H group for its ppp orientational curve that can satisfactorily interpret the free O-H SFG data, a well-established fact in the literature [2,3,5]. Therefore, it is questionable and erroneous to interpret the SFG data with such a broad orientational distribution as in the Ref. [1].

In conclusion, the interpretation of the SFG experimental data for a broad, exponentially decaying distribution of the OH distribution in Ref. [1] is questionable and was based on erroneous analysis of the SFG data. Such interpretation is also inconsistent with the prior SFG examinations based on sound spectral analysis [2-5]. Therefore, the main conclusion that both the SFG experimental data and MD simulation supports broad exponential decaying distribution cannot be reliably established. Future effort should be focused on the reasons behind the discrepancies between the experimental data that suggest a narrow orientational distribution and the simulation results that suggest an otherwise broad exponential decaying distribution, instead of accepting the impression of the never-had agreement between them as concluded in Ref. [1].

Authors thank the support from the National Natural Science Foundation of China (21727802, 21673285), the Ministry of Science and Technology of China (MOST No. 2017YFB0602205), and funding from the Shenzhen city (JCYJ20170307150520453).

Wei Gan
State Key Laboratory of Advanced Welding and Joining, and School of Science, Harbin Institute of Technology (Shenzhen), University Town, Shenzhen, Guangdong, 518055, China

Ran-Ran Feng
Key Laboratory of Microgravity, Beijing Key Laboratory of Engineered Construction and Mechanobiology, Institute of Mechanics, Chinese Academy of Sciences, Beijing 100190, China

Hong-Fei Wang*

Department of Chemistry and Shanghai Key Laboratory of Molecular Catalysis and Innovative Materials, Fudan University, 220 Handan Road, Shanghai 200433, China
wanghongfei@fudan.edu.cn